\documentclass[aps,twocolumn,preprintnumbers,amsmath,amssymb,superscriptaddress,prl,10pt]{revtex4-1}
\usepackage{hyperref}
\usepackage{amsmath}
\usepackage{graphicx}
\usepackage{amssymb}
\usepackage{color}
\usepackage[version=3]{mhchem}
\usepackage{siunitx}

\newcommand{\micron}{\si{\micro\meter}}
\newcommand{\nitride}{\ce{Si3N4}}

\begin{document}

\title{High-Performance Silicon-Based Multiple Wavelength Source}

\author{Jacob S. Levy}
\email{jsl77@cornell.edu}
\affiliation{School of Electrical and Computer Engineering, Cornell University, Ithaca, NY 14853}
\author{Kasturi Saha}
\affiliation{School of Applied and Engineering Physics, Cornell University, Ithaca, NY 14853}
\author{Yoshitomo Okawachi}
\affiliation{School of Applied and Engineering Physics, Cornell University, Ithaca, NY 14853}
\author{Mark A. Foster}
\affiliation{Department of Electrical and Computer Engineering, The Johns Hopkins University, Baltimore, MD 21218}
\author{Alexander L. Gaeta}
\affiliation{School of Applied and Engineering Physics, Cornell University, Ithaca, NY 14853}
\author{Michal Lipson}
\affiliation{School of Electrical and Computer Engineering, Cornell University, Ithaca, NY 14853}
\affiliation{Kavli Institue at Cornell for Nanoscale Science, Cornell University, Ithaca, NY 14853}

\begin{abstract}
We demonstrate a stable CMOS-compatible on-chip multiple-wavelength source by filtering and modulating individual lines from a frequency comb generated by a microring resonator optical parametric oscillator.. We show comb operation in a low-noise state that is stable and usable for many hours. Bit-error rate measurements demonstrate negligible power penalty from six independent frequencies when compared to a tunable diode laser baseline. Open eye diagrams confirm the fidelity of the 10 Gb/s data transmitted at the comb frequencies and the suitability of this device for use as a fully integrated silicon-based WDM source. \\
\end{abstract}

\maketitle

%introduction
\noindent Silicon-based integrated photonics aims to deliver on-chip optical communications 
networks with bandwidths orders of magnitude larger than electronic networks. As the microelectronics industry moves to multi-core and multi-processor chips, CMOS-compatible photonics will replace much of the electronic communications backbone. A key benefit of optical communication systems is wavelength-division multiplexing (WDM) which enables a single waveguide to carry multiple data streams and is essential to reach the full bandwidth potential of photonic integrated circuits. Many components necessary for on-chip optical interconnects such as filters~\cite{Popovic2006}, modulators~\cite{Xu2005}, switches~\cite{Li2008} and detectors~\cite{Ahn2007} have been demonstrated over the past decade. However, an integrated silicon on-chip source capable of generating the many wavelengths necessary to drive the network has been elusive. Because silicon is an indirect band gap material, approaches thus far have focused on integrating III-V active devices by bonding, such as microdisk lasers~\cite{VanCampenhout2007} or a hybrid waveguide~\cite{Fang2006}. Although both methods can be replicated on-chip to generate multiple wavelengths, scaling to the hundreds of wavelengths envisioned by optical network architectures~\cite{Lee2008,Joshi2009} quickly becomes power hungry and space consuming. All-optical approaches have included utilizing the Raman effect in silicon~\cite{Boyraz2004}. Although this process can be cascaded~\cite{Rong2008}, the wavelength separation is determined by the Raman shift which is inadequate for WDM standards. 

%what we did before
In contrast, we recently demonstrated a fully integrated CMOS-compatible optical parametric oscillator (OPO)~\cite{Levy2010} using a silicon nitride microring resonator. The device produces a comb of wavelengths evenly spaced in frequency, as determined by the resonator free spectral range (FSR), with a single pump input. This source could readily be used as an on-chip multiple wavelength source matched to the telecom WDM standards.

\begin{figure}[!b]
\centerline{\includegraphics[width=\columnwidth]{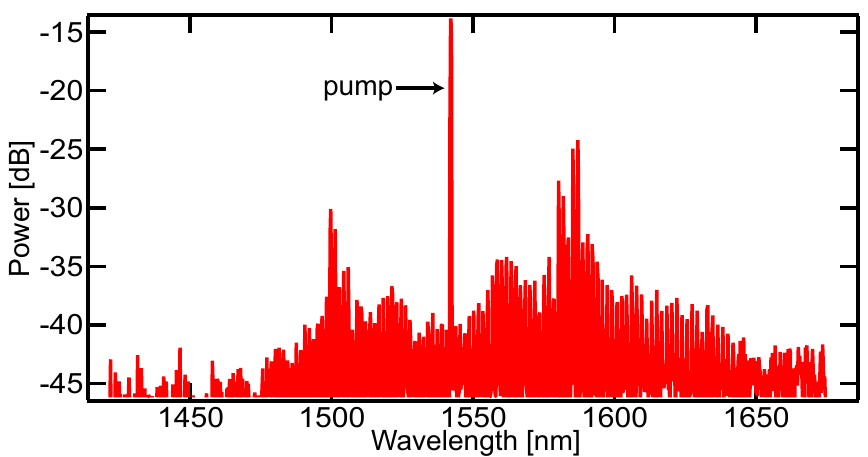}}
  \caption{
  Optical spectrum of the generated frequency comb at the output of the \nitride{} waveguide with a pump wavelength of 1541 nm. The comb is in a stable low-noise state with this spectrum and we are able to manipulate and process the generated frequencies when filtering out the pump. }
\label{spectrum}
\end{figure}

%why microresonator combs important/this work
In this work, we examine the characteristics of the generated comb lines of the \nitride{} OPO and analyze their fidelity as WDM sources. In principle, the highly nonlinear process used to generate the comb could induce signal and noise. The parametric process which generates the new wavelengths also induces a distinct phase relationship between the generated comb lines~\cite{Kippenberg2011,Foster2011}. Additionally, the large quality factor, \textit{Q}, of the resonator leads to high circulating powers. The large intensities could also induce instability in the system due to the sensitivities of both the parametric gain and cavity lineshape to power and temperature fluctuations. Therefore, a robust measurement of the generated frequencies is required to demonstrate the usefulness of the on-chip source. Note that previous works have shown comb generation using parametric gain in microresonators~\cite{Del'Haye2007,Grudinin2009,Levy2010,Razzari2010}, but an analysis of the fidelity and stability of encoded data on filtered individual comb lines has not yet been examined.

%comb generation
The bit-error rate (BER) measurements are performed on a 116-~\micron{} radius \nitride{} ring resonator OPO that generates over 100 new wavelengths across a 200 nm span (Figure~\ref{spectrum}) by pumping the cavity at a resonance near 1541 nm. The intrinsic \textit{Q} of the ring is on the order of $10^6$ which induces a cavity enhancement of the input pump by a factor 200. Utilizing the nonlinearity of the silicon nitride, signal and idler wavelengths initially oscillate at either side of the pump at a threshold power level at which the parametric gain induced by the pump is greater than the cavity loss at some neighboring resonance. After threshold, \SI{50}{\milli\watt} ~\cite{Levy2010}, with increased pump power cascaded four-wave mixing generates new frequencies at each cavity resonance. Recently, we have shown that the spacings between these frequencies are even to within 1 part in $10^{15}$ of the center frequency~\cite{Foster2011}. In our previous studies, we have shown spectrally flat and very broad frequency combs spanning 2/3 of an octave. The broadband combs can exhibit a high level of intensity noise measured with an RF spectrum analyzer indicating that the generated wavelengths are unstable in frequency and amplitude. %{michal wants to remove} For use as an on-chip source, the generated wavelengths must lack frequency or power drift or the need to re-initialize oscillation in the microring. 

 %experimental set-up
We achieve stable operation of the comb by tuning the pump wavelength to a spectral point in the microring resonance where the comb intensity noise drops. We tune the pump laser into resonance from the "`blue-detuned"' side to avoid thermal bistability. As we tune into the resonance the RF intensity noise drops by more than 25 dB and measurement of the temporal output suggests the comb is mode-locked~\cite{Foster2011}. In this low-noise operation state, a stable comb can be maintained for hours. 

We filter and modulate individual wavelengths generated in the ring resonator and measure the BER and power penalty at 10-Gb/s. The output of the chip is collected and sent through a bandpass filter to isolate a 10-nm spectral range consisting of six newly generated wavelengths. The power at each wavelength coupled from the ring to the bus waveguide is on the order of \SI{1}{\milli\watt}; however, the inefficiency of our off-chip collection and filtering require amplification for further processing. Therefore, we use a low-noise erbium-doped fiber amplifier (EDFA) to increase the power of the filtered comb lines. We then use a \SI{1}{\nano\metre} tunable filter to select individual wavelengths. We modulate the selected wavelength with a lithium niobate Mach-Zehnder modulator driven by a pattern generator to imprint a $2^{31} - 1$ non-return-to-zero psuedo-random bit sequence (PRBS) at a data rate of 10 Gb/s. The modulated signal is then sent to either a sampling oscilloscope to generate eye diagrams or to a variable optical attenuator (VOA), then into a 10 Gb/s lightwave receiver and a bit-error-rate tester (BERT). For the baseline measurement of our set-up, we modulate a tunable external cavity diode laser, operating at a wavelength near the filtered comb lines.

\begin{figure}[!b]
\centerline{\includegraphics[width=\columnwidth]{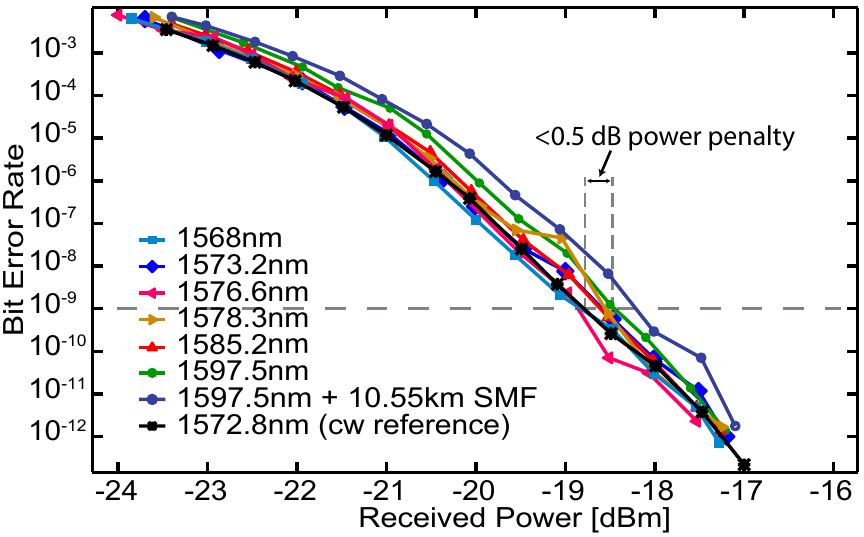}}
  \caption{
  BERT measurements of the filtered comb lines. The cw reference measurement acts as a back-to-back baseline with which to compare the modulated comb lines. We note error-free operation of the comb lines and a minimal power penalty measured at a bit-error-rate of $10^{-9}$}
\label{BER}
\end{figure}

%power penalty in BER
In the low-noise mode-locked state, we measure negligible power penalty of the filtered comb lines demonstrating suitability of the source for an on-chip WDM network. The BER as a function of received optical power is plotted in Figure~\ref{BER}. We observe a negligible power penalty for the tested comb lines as compared to the baseline across a 30-~\si{\nano\meter} span, where power penalty is defined as the difference in received power at $10^{-9}$ errors between the baseline and test data. Additionally, we achieve error-free operation, that is a BER less than $10^{-12}$ for the generated wavelengths. Although we expect the BER to be linear with received power, we observe a slight curvature in the data. Since the curvature is the same for the baseline and data measurement we conclude it originates from our test equipment and is not fundamental to the microring wavelength generator. Additionally, we transmit data from one of the comb lines through a 10 km spool of optical fiber to show suitability for long-haul optical communications. In this case, we incur a very small power-penalty in reference to the same comb line without the added fiber. 

\begin{figure}[htb]%
\includegraphics[width=\columnwidth]{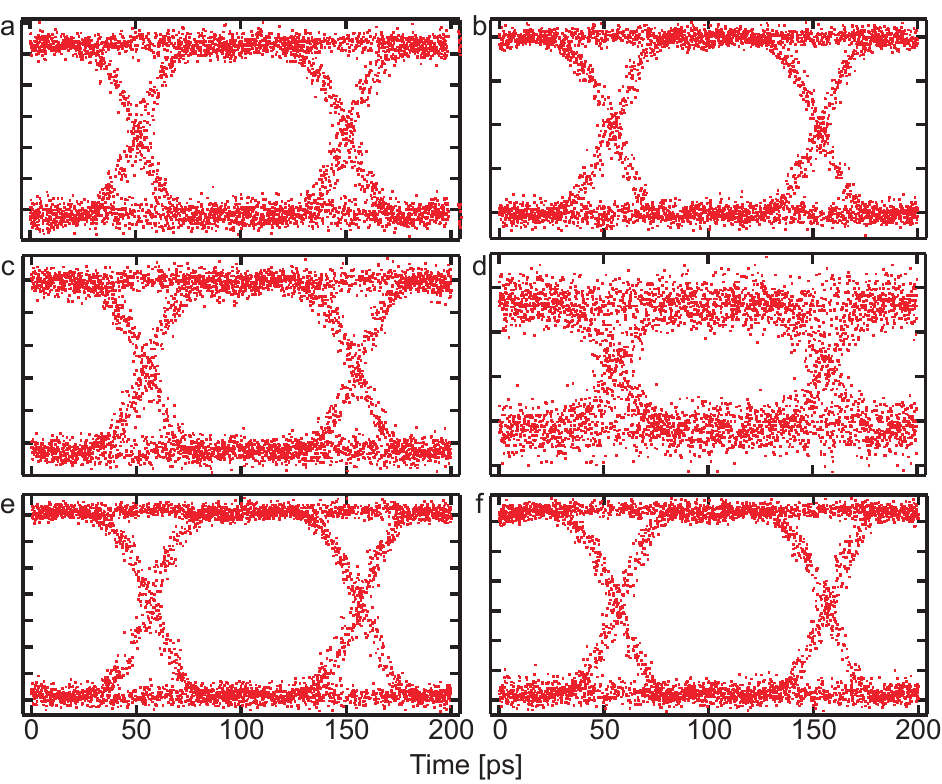}%
\caption{ Eye diagrams for the six measured comb lines generated by the microcavity for (a) 1568 nm (b) 1573.2 nm (c) 1576.6 nm (d) 1578.3 nm (e) 1585.2 nm and (f) 1597.5 nm. The clean and open eye confirms the BER data showing no power-penalty.
}%
\label{EYEZ}%
\end{figure}

%eye diagram data
Open eye diagrams confirm the low power penalty of the modulated spectral lines. Figure~\ref{EYEZ} shows the eye diagram for each comb line on which we performed BER measurements. As expected, the 10 Gb/s signal shows up clearly on the oscilliscope with minimal noise and no closing of the pattern eye. Therefore, we have independently filtered and modulated this comb line without distortion from generated neighbors.

%summary conclusions
In summary, we characterize the performance of the multiple wavelength source generated from a microring resonator OPO and show that it is suitable for an on-chip optical communications network. The frequency comb used in this experiment had a line spacing of ~200 GHz. We can tune this parameter by modifying the resonator's radius, which allows for great flexibility in channel spacing and the potential to meet standard WDM specifications. Although in this work filtering and modulation are done off-chip, through the incorporation of cascaded silicon electro-optic ring resonators~\cite{Manipatruni2010}, the same functionality can be performed on-chip for a completely integrated system. In that case, each silicon ring modulator would be aligned to one of the hundreds of generated wavelengths to enable on-chip data transmission rates of over 1 Tb/s. \\

%acknowledgements
The authors acknowledge DARPA for supporting this work under the Optical Arbitrary Waveform Generation Program and the MTO POPS Program and by the Center for Nanoscale Systems, supported by the National Science Foundation and the New York State Office of Science, Technology and Academic Research.   This work was performed in part at the Cornell NanoScale Facility, a member of the National Nanotechnology Infrastructure Network, which is supported by the National Science Foundation (Grant ECS-0335765).

\bibliography{C:/Users/Jacob/Documents/library}
\bibliographystyle{ol}
\end{document}